\documentclass[sigconf]{acmart}

\usepackage{amsmath}
\usepackage{graphicx}
\usepackage{multirow}
\usepackage{verbatim}
\AtBeginDocument{%
  \providecommand\BibTeX{{%
    \normalfont B\kern-0.5em{\scshape i\kern-0.25em b}\kern-0.8em\TeX}}}

\setcopyright{acmcopyright}
\copyrightyear{2018}
\acmYear{2018}
\acmDOI{XXXXXXX.XXXXXXX}

\acmConference[Conference acronym 'XX]{Make sure to enter the correct
  conference title from your rights confirmation emai}{June 03--05,
  2018}{Woodstock, NY}
\acmBooktitle{Woodstock '18: ACM Symposium on Neural Gaze Detection,
 June 03--05, 2018, Woodstock, NY} 
\acmPrice{15.00}
\acmISBN{978-1-4503-XXXX-X/18/06}

\begin{document}

\title{Unsupervised Human Activity Recognition through \\ Two-stage Prompting with ChatGPT}

\author{Qingxin Xia}
\affiliation{%
  \institution{Graduate School of Information Science and Technology, Osaka University}
\city{Osaka}
\country{Japan}
}
\email{xia.qingxin@ist.osaka-u.ac.jp}

\author{Takuya Maekawa}
\affiliation{%
  \institution{Graduate School of Information Science and Technology, Osaka University}
  \city{Osaka}
  \country{Japan}
}
\email{maekawa@ist.osaka-u.ac.jp}

\author{Takahiro Hara}
\affiliation{%
  \institution{Graduate School of Information Science and Technology, Osaka University}
\city{Osaka}
\country{Japan}
}
\email{hara@ist.osaka-u.ac.jp}

\begin{abstract}

Wearable sensor devices, which offer the advantage of recording daily objects used by a person while performing an activity, enable the feasibility of unsupervised Human Activity Recognition (HAR).
Unfortunately, previous unsupervised approaches using the usage sequence of objects usually require a proper description of activities manually prepared by humans.
\textcolor{black}{Instead, we leverage the knowledge embedded in a Large Language Model (LLM) of ChatGPT.}
Because the sequence of objects robustly characterizes the activity identity, it is possible that ChatGPT already learned the association between activities and objects from existing contexts.  
However, previous prompt engineering for ChatGPT exhibits limited generalization ability when dealing with a list of words (i.e., sequence of objects) due to the similar weighting assigned to each word in the list.
In this study, we propose a two-stage prompt engineering,
which first guides ChatGPT to generate activity descriptions associated with objects while emphasizing important objects for distinguishing similar activities; 
then outputs activity classes and explanations for \textcolor{black}{enhancing the contexts that are helpful for HAR.}
To the best of our knowledge, this is the first study that utilizes \textcolor{black}{ChatGPT} to recognize activities using objects in an unsupervised manner.
We conducted our approach on three datasets and demonstrated the state-of-the-art performance.

\end{abstract}



\received{20 February 2007}
\received[revised]{12 March 2009}
\received[accepted]{5 June 2009}

\maketitle

\section{Introduction}
Human Activity Recognition (HAR) using wearable sensor data regarding body movements and interactions with daily objects has been adopted in many applications such as daily life \cite{PersonalizeHAR}, smart factory \cite{ Jaime, yoshimura}, and nursing care \cite{inoue}.
However, many predominant methods for HAR require ground truth labels, which are expensive to obtain due to a large amount of human effort, expertise, and time consumption.

%
Unsupervised learning techniques have gained increasing attention in recent years. By leveraging relevant information about the activities \cite{maekawa16, docxia}, the unsupervised learning approach can reach state-of-the-art performance on par with supervised. 
For example, prior studies \cite{Philipose2004, Wyatt2005} proposed to model activities as sequences of object use and realized the possibility of discriminating between activities by taking as features the objects used. These studies employ RFID wristbands on users' wrists to automatically record objects and thus recognize activities without using activity labels for training.
\textcolor{black}{However, the description of activities in these studies should be prepared manually by humans, and the performance of these studies is heavily dependent on the quality of these descriptions.}

\textcolor{black}{Recently, large language models have presented a remarkable performance on natural language processing tasks, one of the most famous applications is ChatGPT \cite{gpt4}.}
Because a large corpus of text data was used to pre-train ChatGPT, it may contain various aspects of information about the activities, including the association between the activities and objects used, e.g., the object name ``trash can'' is frequently mentioned when describing the cleanup activity.
Leveraging proper prompt engineering, ChatGPT can be effectively utilized to automatically generate activity descriptions for HAR by considering the associated objects.
However, existing prompt engineering approaches \cite{promptsurvey} may not yield good activity classification results, especially for those activities that share overlapping objects, since the object names are assigned a similar weighting in the prompt.
Therefore, designing prompt engineering to emphasize the important objects for differentiating between activities is necessary.


In this study, we present a novel two-stage prompt engineering approach for ChatGPT to generate activity descriptions based on sequences of objects and output activity classification results in an unsupervised manner. 
The first stage is knowledge generation, which generates descriptions of activities containing object information by employing ChatGPT to distinguish between two activities. 
Because we generate activity descriptions specifically tailored for activity differentiation purposes, the context generated by ChatGPT during this process may contain valuable information that can be utilized for activity classification. 
The second stage is answer generation, which utilizes the knowledge generated from the first stage in conjunction with knowledge prompt engineering \cite{knowledgePrompt} to \textcolor{black}{output the prediction of activities}.
Knowledge prompt engineering requests an explanation for the HAR result based on the knowledge in the prompt. We utilize it to enhance the \textcolor{black}{contexts in the prompt} that are helpful for HAR.
By integrating the two-stage prompt engineering, we aim to guide ChatGPT to automatically focus on important contexts for differentiating between activities, thus improving the model's HAR performance.

The contributions of this study are listed as follows:
\begin{enumerate}
    \item We proposed an unsupervised HAR approach using the sequence of objects. To the best of our knowledge, this is the first to utilize ChatGPT for predicting activities based on the usage of objects.
    \item We proposed a two-stage prompt engineering that promotes ChatGPT to differentiate activities via objects.
    \item We compare our approach to three prompt engineering baselines. Our approach demonstrates the best performance on three HAR benchmark datasets.
\end{enumerate}

\section{Related Work}

\begin{figure*}[htbp]
\centering
\includegraphics[width=17cm]{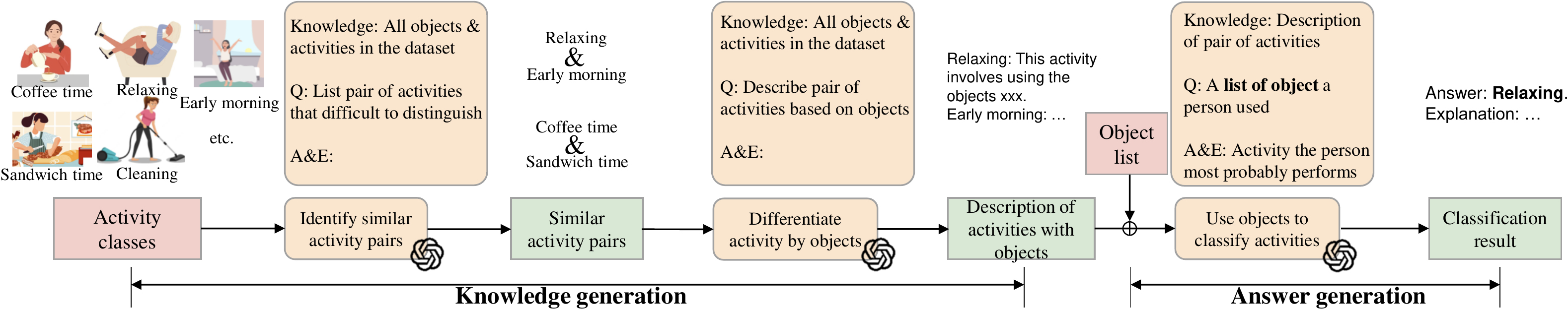}
\caption{The overview of the Proposed approach. The pink background rectangles show the input of the model, the orange rectangles show the procedures done by ChatGPT, and the green rectangles show the output results of each prompt for ChatGPT processing.}
\Description{TBD.}
\label{fig:overview}
\end{figure*}



\textbf{HAR using Wearables}.
HAR plays an important role in monitoring human behavior, which is helpful for understanding individuals' wellness and supporting personalized systems.
HAR using wearable devices \cite{wearablereview} has become one of the most important tasks in the ubiquitous community due to the cost-effectiveness of wearables compared to other sensors like video cameras, as well as their ability to collect activity data without excessively compromising privacy unrelated to HAR.
There are many applications that use data collected from wearables for HAR.
For example, by recognizing the workers' activities performed in a logistics center, the bottleneck activities could be identified \cite{Jaime}.
Francisco et al. \cite{complexact} utilized a hybrid model with both convolutional and recurrent structure layers to recognize human activities in daily life.
Chen et al. \cite{complexresidual} employed a residual neural network architecture to extract features from sensor data and recognize activities.
These studies can be helpful for developing individual assistant systems employed in smart homes and etc.
However, in real settings, activities tend to be complex and consist of multiple actions. Training HAR models for more complex activity classes typically requires a larger number of activity labels, which can be impractical.

\textbf{Unsupervised Activity Recognition}.
To reduce the number of labels used for training, many unsupervised learning techniques have been attracting attention.
Maekawa et al. \cite{maekawa16} proposed an unsupervised method for identifying each iteration of assembly work by utilizing acceleration data collected from the workers' wrists.
Xia et al. \cite{docxia} built upon the aforementioned study for estimating the starting and ending times of each activity within the assembly work by leveraging the information derived from process instructions.
Hiremath et al. \cite{smarthome} employed a cluster-based approach that utilized multiple sensors in a smart home setting to recognize human activities.
However, the approaches proposed by these studies strongly rely on the characteristics of sensor data and the environmental settings, limiting their applicability in broader scenarios.

\textbf{HAR Relying on Usage of Objects}.
Owning to the above challenges, prior studies such as Philipose et al. \cite{objectphilipose} and Tapia et al. \cite{tapia2004activity} have suggested discriminating between many activities by using the objects that the user used because the objects strongly correlate to that of natural-language instructions (e.g., recipes).
Wyatt et al. \cite{unsupervisedobject} tried to model the activity of daily life using the object sequence with object labels recorded by the wristband.
Patterson et al. \cite{objectpatterson} proposed an approach to infer daily life activities from the aggregated object instance.
%
\textcolor{black}{However, the aforementioned works rely on having explicit descriptions of activities, and the quality of these descriptions directly affects the performance of the activity recognition models. }

In this study, we propose an unsupervised learning approach to recognize human activities using the object sequence that the user used for each activity. The description of activities will be automatically generated through ChatGPT.


\section{Methodology}

As shown in Figure \ref{fig:overview},
given a list of objects that the user interacted with, this study employs a two-stage prompt engineering to automatically generate knowledge about activities and then recognize activities using the object sequence in an unsupervised manner.

\subsection{Problem Setup}

\textcolor{black}{Let $L$ represent a sequence of objects as input, }
$K_p$ represents the knowledge regarding $L$, $A$ represents the set of answers including the activity classes, and $pG(\cdot)$ is the pre-trained language model in ChatGPT.
The model aims to predict an activity $a \in A$, which is formulated as follows:
\begin{align}
\hat{a} = \underset{a\in A}{\text{arg max }} pG(a|L, K)
\label{formula:prompt}
\end{align}

\subsection{Knowledge Generation}
Because we did not manage to fine-tune the LLM for our specific task, 
the knowledge embedded in the prompt becomes crucial for enabling ChatGPT to generate precise activity recognition results. 
Merely incorporating the sequence of object names in the prompt is challenging to distinguish between similar activities that involve overlapping objects. Therefore, additional knowledge needs to be incorporated into the prompt to guide ChatGPT in focusing on important objects.
In this study, we automatically generate the additional knowledge $K$ using the pre-trained language model in ChatGPT. 

As shown in Figure \ref{fig:overview}, this structure consists of two prompts. The first prompt \textcolor{black}{is designed to identify} several pairs of activities that are difficult to distinguish from objects of usage. Then, utilizing the outputs from the first prompt, the second prompt outputs the description of each activity using objects.

In the first prompt, we aim to \textcolor{black}{generate} $k$ pairs of activities that are difficult to distinguish from each other using the implicit knowledge of ChatGPT. 
Let $O$ represent all the object names in the dataset, and $q_p$ represent the question in this prompt. The pairs of activities generated by the first prompt are described as follows:
\begin{align}
E =  \{e_p: e_p \sim pG(e_p|q_p, A, O), p=1,2,\dots,k\},
\label{formula:pair}
\end{align}
where $e_p$ is a pair of activities expressed in the textual form of "activity A and activity B."
According to the formula, the answer will be generated based on the object names, activity classes in the prompt, and the knowledge embedded in $pG(\cdot)$.
The implicit knowledge embedded in the $pG(\cdot)$ can be helpful to identify similar activities based on the knowledge provided in the prompt.
The details of the prompt are composed of All objects, All activities, Question, and Explain and Answer. We append the following sentences for each component:
\textit{All objects in the dataset: \{All object names\};}
\textit{All activities in the dataset: \{All activity classes\};}
\textit{Question: List $k$ pair of activities in [All activities] that is difficult to distinguish;}
\textit{Answer and Explanation: }.
In the sentences, \{$\cdot$\} are placeholders that are uniformly replaced with the corresponding text before being fed into the LLM.

The objective of the second prompt is to generate descriptions of activities $K$ using the knowledge $E$ obtained from the first prompt.
For each $e_p \in E$, we generate the descriptions for the pair of activities $k_p \in K$ simultaneously. By asking ChatGPT to differentiate the pair of activities in the prompt, the important objects in differentiating the activities can be obtained in the output sentence. 
the idea of generating $k_p$ for each $e_p$ is formulated as follows: 
\begin{align}
\hat{k_p} = \underset{k_p \in K}{\text{arg max }} pG(k_p|e_p, A, O), p=1,2,\dots,k\
\end{align}

The details of the prompt are similar to the first prompt with a different Question. The sentence is organized as follows:
\textit{Question: Differentiate \{$e_p$\} activities based on  objects.}
We replace the placeholder for each corresponding text of $e_p$ before fed into the LLM, and then combine every output to get $K$.

Here is an example $k_p$ from the prompt:
\textit{Cleanup: During the Cleanup activity, the objects used should be put back to their original place or to the dishwasher. This includes objects such as the Bread Cutter, Knifes, Plates, Glass, Cup, and Plate. $\backslash n$     
Early Morning: During the Early Morning activity, the objects used can include the Switch, Drawer3 (lower), Drawer2 (middle), Drawer1 (top), Fridge, and Lazychair. These objects are used to perform various activities such as turning on the lights, opening drawers, and getting out of bed.
}

\subsection{Answer Generation}

According to Knowledge Generation, ChatGPT automatically generates the descriptions of activities associated with objects.
In this part, we discuss predicting activities from a sequence of object names via generated knowledge and ChatGPT.
In this step, the Knowledge component in the prompt engineering becomes the descriptions of activities $K$ generated from the first stage.
Since the activity descriptions in the prompt are already generated for differentiating similar activities, when a new sequence of objects is sent to the prompt, ChatGPT can easily focus on the important objects based on $K$.
Besides, we utilize knowledge prompt engineering to improve ChatGPT's ability for HAR. By requesting explanations for HAR based on the prompt's knowledge, we \textcolor{black}{emphasize the important contexts (e.g., object names) in the prompt} used for activity recognition. 
The details of the prompt are displayed as follows:
\textit{All activities in the dataset: \{All activity classes\};}
\textit{Name and Description of activities: \{$K$\};}
\textit{Question: A list of objects a person used that ordered in time: \{sequence of object names\}. Output the name of the activity the person most probably performs;}
\textit{Answer and Explanation: }.

Here is an example output by the prompt:
\textit{Answer: Early morning.$\backslash n$ Explanation: The objects used in this list suggest that the person is most likely performing an early morning activity, such as turning on the lights, opening drawers, and getting out of bed.}

\begin{table*}[]
\caption{An overview of the three datasets. }
\begin{tabular}{l|l|l} \toprule
 &
  \textbf{Activity Classes} &
  \textbf{Object Names} \\ \hline
\textbf{Opportunity} &
  \begin{tabular}[c]{@{}l@{}}Relaxing, coffee time, sandwich time, cleanup,\\ early morning\end{tabular} &
  \begin{tabular}[c]{@{}l@{}}Salami, milk, fridge, bottle, glass, dishwasher, salami knife, \\ spoon, cup, bread, plate, drawer2 (middle), drawer3 (lower),\\  door1, door2, table\end{tabular} \\ \hline
\textbf{50Salads} &
  \begin{tabular}[c]{@{}l@{}}Cut\_and\_mix\_ingredients, prepare\_dressing,\\ serve\_salad\end{tabular} &
  \begin{tabular}[c]{@{}l@{}}Cucumber, tomato, cheese, bowl, lettuce, ingredients, oil, \\ vinegar, salt, pepper, dressing, plate\end{tabular} \\ \hline
\textbf{CMU-MMAC} &
  Cook: Brownie, egg, pizza, salad, sandwich &
  \begin{tabular}[c]{@{}l@{}}Fork, fridge, bread, egg, sink, big\_bowl, cup, frying\_pan, \\ brownie\_box, scissors, brownie\_bag, baking\_pan, plate, \\ oven, small\_bowl, drawer, oil, grater, cheese, pepper, knife, \\ sausage, cucumber, caesar\_dressing, carrot, broccoli, celery \end{tabular} \\ \bottomrule
\end{tabular}
\label{tab:dataset}
\end{table*}

\section{Experimental Evaluation}
\subsection{Dataset}

\begin{table}[htbp]
\caption{The F1-score (\%) of three datasets.}
\label{tab:accuracy}
\small
\begin{tabular}{c|ccc}
\toprule
\multicolumn{1}{l|}{}                                               & \textbf{Opportunity} & \textbf{50Salads} & \textbf{CMU-MMAC} \\ \hline
Zero-shot                                                           & 53.61                & 64.47             & 76.32                 \\ \hline
\begin{tabular}[c]{@{}c@{}}Retrieval-based\\ knowledge\end{tabular} & 48.19                & 93.87             & 71.05                 \\ \hline
Few-shot                                                            & 73.08                & 90.61             & 100.00                 \\ \hline
\textbf{Proposed}                                                   & 91.15                & 100.00              & 100.00                 \\ \bottomrule
\end{tabular}
\end{table}

We evaluate our method on three datasets. 
An overview of the three datasets is displayed in Table \ref{tab:dataset}.

\noindent \textbf{Opportunity dataset} \cite{opportunity}.
This is a widely used benchmark dataset for HAR from wearables, which records a variety of activities of daily life, with five classes of activities.
When the user interacts with an object, the object label will be recorded.

\noindent \textbf{50 Salads} \cite{50salads}.
This dataset collects accelerometer data in a kitchen room. Three classes of activities are recorded, corresponding to different states while cooking.
Object labels are collected when the user is using the objects, such as the knife, glass, etc.

\noindent \textbf{CMU-MMAC} \cite{cmummac}.
This is a multi-modal database that contains multi-modal measures of the human activity of subjects performing the tasks involved in cooking and food preparation. Twenty-five subjects have been recorded cooking five recipes with the object they used during experiments.

\textcolor{black}{In these datasets, each pre-segmented time period corresponds to a complete activity, and the object usage at each timestamp is recorded and converted into the corresponding object name.}

\subsection{Result}

\begin{figure*}[htbp]
\centering
\includegraphics[width=17cm]{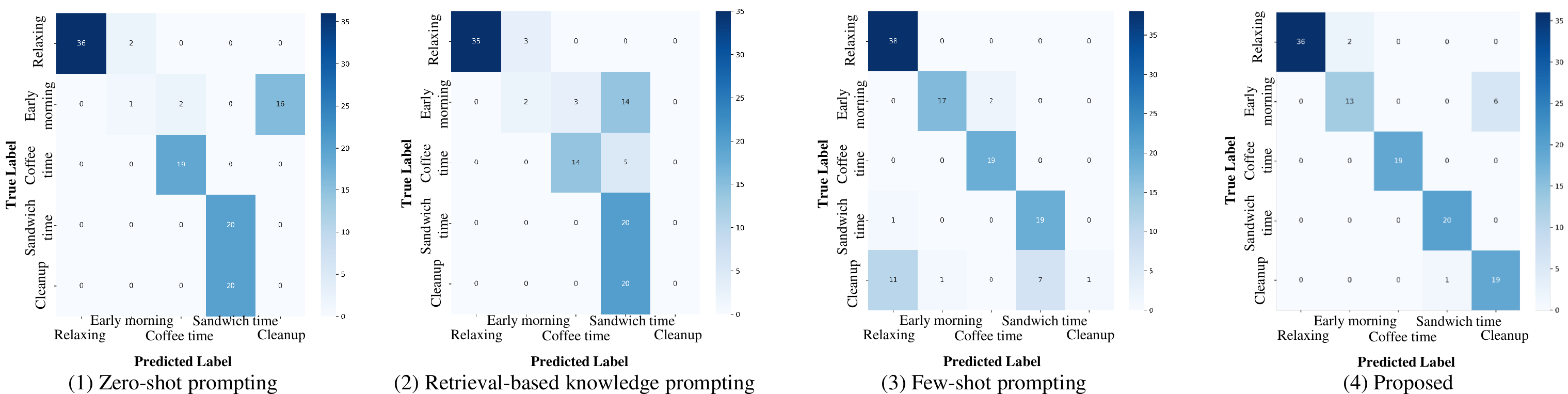}
\caption{The confusion matrix of comparative prompting methods and the proposed method of Opportunity dataset.}
\Description{TBD.}
\label{fig:exampleresult}
\end{figure*}


We conducted a comparative analysis of our proposed prompt engineering approach against three widely used techniques. 


\noindent \textbf{Zero-shot prompting} \cite{zeroshot}: This approach does not contain any knowledge statement in the prompt. The prompt only consists of all object names and classes of activities.

\noindent \textbf{Retrieval-based knowledge prompting} \cite{knowledgePrompt}: In addition to the object names and activities, this approach utilizes the knowledge statement retrieved from the dataset or the sentences in appropriate sources, such as Wikipedia.

\noindent \textbf{Few-shot prompting} \cite{fewshot}: In addition to the zero-shot prompting, this approach employs some examples in the knowledge statement. In this study, for each class of activity, we provide an object name sequence and the corresponding activity as question and answer.
Unlike the other approaches, this approach requires costs for preparing the knowledge statement. 

For a fair comparison, we implemented prompting engineering using the same parameters. We use the ``text-DaVinci-003'' of the GPT-3.5 family as the knowledge generator, the temperature was set to 0, and the top\_$p$ was set to 0.5.

\textcolor{black}{We used the sequence of objects for each segmented time period as input and output for the predicted activity class via different prompt engineering.
We evaluated the model performance using the micro average F1-score.}

In Table \ref{tab:accuracy}, our proposed approach outperformed the baselines among all datasets.
In these datasets, objects overlap between activities. According to zero-shot prompting, if only a sequence of objects was provided, it is insufficient to distinguish between activities.
Retrieval-based knowledge prompting performed poorly on the Opportunity dataset, while the proposed approach always has good performances on various datasets. This result suggested that the knowledge generated by the proposed approach is more robust than the knowledge prepared by retrieval-based knowledge prompting.
Few-shot prompting, \textcolor{black}{which involved some examples manually prepared by humans in the prompt}, demonstrated the highest performance among the three baselines. However, it did not perform as effectively as the proposed approach on the Opportunity dataset. This discrepancy can be attributed to the greater flexibility of objects used in the Opportunity dataset compared to the cook-related datasets (i.e., 50Salads and CMU-MMAC). 
The results imply that the proposed prompt engineering has the ability to generate \textcolor{black}{high-quality} knowledge based on the existing contexts within ChatGPT in an unsupervised manner. 

Figure \ref{fig:exampleresult} presents the confusion matrix of the Opportunity dataset using different prompting engineering. In the dataset, the cleanup activity contains many objects in the other activities. By generating descriptions of activities from pair of similar activities, the proposed prompting approach successfully identified cleanup activity from other activities.


\section{Conclusion}

In this paper, we propose a two-stage prompting engineering used for ChatGPT to recognize activities using objects the person used in an unsupervised manner.
Our proposed prompt engineering is able to generate descriptions of activities for various datasets automatically.



\bibliographystyle{ACM-Reference-Format}
\bibliography{reference}

\end{document}